\documentclass[aps,reprint,showpacs,preprintnumbers,amsmath,amssymb,prl,superscriptaddress]{revtex4-1}

\usepackage{graphicx}
\usepackage{dcolumn}
\usepackage{bm}
\usepackage{multirow}%
\usepackage[utf8]{inputenc}
\usepackage[T1]{fontenc}
\usepackage{longtable}

\usepackage{verbatim}
\usepackage{color}

\newcommand{\bt}{BaTiO$_3$}
\newcommand{\nn}{NaNbO$_3$}
\newcommand{\kn}{KNbO$_3$}
\newcommand{\st}{SrTiO$_3$}
\newcommand{\ct}{CaTiO$_3$}
\newcommand{\kt}{KTaO$_3$}
\newcommand{\nt}{NaTaO$_3$}

\newcommand{\la}{LaAlO$_3$}

\begin{document}

\title{$d^0$ Ferromagnetic Interface Between Non-magnetic Perovskites}
\author{R. Oja}
\affiliation{COMP Centre of Excellence, Department of Applied Physics, Aalto University, P.O. Box 11100, 00076 Aalto, Helsinki, Finland}
\author{M. Tyunina}
\affiliation{Microelectronics and Materials Physics Laboratories, University of Oulu, P.O. Box 4500, 90014 Oulun yliopisto, Finland}
\affiliation{Institute of Physics, Academy of Sciences of the Czech Republic, Na Slovance 2, 182 21 Prague 8, Czech Republic}
\author{L. Yao}
\affiliation{NanoSpin, Department of Applied Physics, Aalto University, P.O. Box 15100, 00076 Aalto, Helsinki, Finland}

\author{T. Pinomaa}
\affiliation{COMP Centre of Excellence, Department of Applied Physics, Aalto University, P.O. Box 11100, 00076 Aalto, Helsinki, Finland}
\author{T. Kocourek}
\author{A. Dejneka}
\author{O. Stupakov}
\author{A. Jelinek}
\author{V. Trepakov}
\affiliation{Institute of Physics, Academy of Sciences of the Czech Republic, Na Slovance 2, 182 21 Prague 8, Czech Republic}
\author{S. van Dijken}
\affiliation{NanoSpin, Department of Applied Physics, Aalto University, P.O. Box 15100, 00076 Aalto, Helsinki, Finland}
\author{R. M. Nieminen}
\affiliation{COMP Centre of Excellence, Department of Applied Physics, Aalto University, P.O. Box 11100, 00076 Aalto, Helsinki, Finland}

\begin{abstract}
We use computational and experimental methods to study $d^0$ ferromagnetism at a charge-imbalanced interface between two perovskites. In SrTiO$_3$/KTaO$_3$ superlattice calculations, the charge imbalance introduces holes in the SrTiO$_3$ layer, inducing a $d^0$ ferromagnetic half-metallic 2D electron gas at the interface oxygen $2p$ orbitals. The charge imbalance overrides doping by vacancies at realistic concentrations. Varying the constituent materials shows ferromagnetism to be a general property of hole-type $d^0$ perovskite interfaces. Atomically sharp epitaxial $d^0$ SrTiO$_3$/KTaO$_3$, SrTiO$_3$/KNbO$_3$ and SrTiO$_3$/NaNbO$_3$ interfaces are found to exhibit ferromagnetic hysteresis at room temperature. We suggest the behavior is due to high density of states and exchange coupling at the oxygen $t_{1g}$ band in comparison with the more studied $d$ band $t_{2g}$ symmetry electron gas.
\end{abstract}

\maketitle

Conventionally, magnetic ordering in solids is associated with atoms with partly occupied 3$d$ or 5$f$ shells. There, large exchange energy $J$ results in high-spin configuration of individual atoms. Coupling of these magnetic moments depends strongly on the material in question. In the metallic band picture of free electrons, ferromagnetic coupling is favored if the density of states at the Fermi level, $D(E_f)$, is high enough. In this case, the exchange energy gained by spin-polarizing the carriers is larger than their corresponding kinetic energy increase. This is known as the Stoner criterion, $JD(E_f) > 1$.

However, high density of states at Fermi level and high exchange interaction are not restricted to these shells only. Hund's coupling is also large in materials with partly filled $2p$ orbitals. Correspondingly, magnetic effects in first-row elements O, N, C and B are a common experimental observation \cite{*[{Reviews in }] coey2005,*volnianska2010}. More specifically, holes in oxygen $p$ orbitals have been widely predicted to result in magnetic moments \cite{elfimov2002,gallego2005,shein2007,*janicka2008,*wu2010,*gruber2012,fischer2011}, and ferromagnetic (FM) ordering of these holes is obtained if the hole density is high enough \cite{elfimov2002,fischer2011,sanchez2008,peng2009}.

The high density of states at Fermi level is difficult to achieve; usually, $2p$ bandwidths are large, meaning the Stoner criterion is unlikely to be satisfied. However, in perovskite oxides, the HOMO oxygen $t_{1g}$ band is very flat and does not hybridize with any other bands \cite{WolframEllialtioglu2006}. This makes it an excellent candidate for realizing Stoner ferromagnetism, if sufficient hole density is achieved.


Simple charge counting dictates that combining one perovskite with neutral (100) layers and another with charged (100) layers results in an interface with additional electrons or holes (termed \textit{n} type or \textit{p} type interface, respectively), to maintain charge neutrality of the complete structure \cite{bristowe2011}. Such charge-imbalanced interfaces between two perovskites have been widely studied, because an interface with a two-dimensional electron gas based on $d$ orbitals exhibits properties such as metallicity \cite{ohtomo2004}, magnetism \cite{brinkman2007,*luders2009} and superconductivity \cite{reyren2007}. Magnetic ordering of these $t_{2g}$ symmetry orbitals is a textbook example of complex behavior in the so-called complex oxides.

An interface with holes on oxygen $p$ orbitals has been much less studied. The oxygen $2p$ band will accommodate holes only if both perovskites have $d^0$ occupancy in the bulk. One problem with the $d^0$ perovskite $p$ type interface is oxygen vacancies \cite{nakagawa2006,pentcheva2006,park2006}, which donate electrons to compensate for the charge imbalance. No 2D metallicity has been observed in the \st{}/\la{} $p$ type interface, suggested to be either due to vacancies or Mott transition to an insulating state \cite{pentcheva2006}. In hole-doped \bt{}, however, both metallic and insulating $p$ magnetic states have been obtained \cite{gruber2012}.

\begin{figure}[ht!]
\includegraphics[width=.38\textwidth]{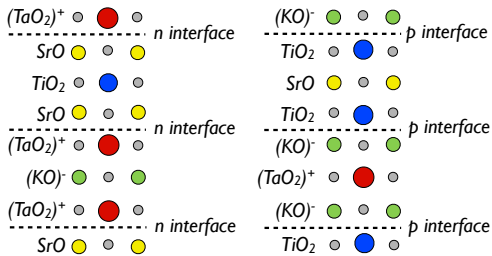}
\caption{\label{fig:superlattice} (Color online) Calculated \st{}/\kt{} \textit{n type} (left) and \textit{p type} (right) superlattices, with nominal ionic charges of the layers shown.}
\end{figure}

In this Letter, we show the $p$ type interface of $d^0$ perovskites \cite{pentcheva2006,oja2009,nazir2011} to be ferromagnetic and half-metallic. As the sample system, we study \st{}/\kt{} superlattices with two similar ($p$ or $n$) interfaces, starting with 1.5 unit cells of both perovskites. This results in a structure with one hole or one electron per supercell compared to nominal ionic charges, Fig. \ref{fig:superlattice}. We compare the results of \textit{ab initio} LSDA+$U$, GGA+$U$ and HSEsol hybrid functional \cite{schimka2011} methods, using the VASP \cite{kresse1993a,*kresse1996b} code and PAW \cite{blochl1994,*kresse1999a} potentials. The GGA employed is PBEsol \cite{perdew2008}. The in-plane lattice constant is fixed to the calculated \st{} lattice constant, to model epitaxy on a \st{} substrate. Ionic relaxations are performed to allow for ionic compensation of the interface dipoles, although their effect is expected to be very small \cite{nazir2011}. 

Considering two-body Coulomb repulsion $U$ or nonlocal exchange (by HSEsol) is necessary due to the failure of LSDA or GGA to properly predict the behavior of partly occupied oxygen $p$ or Ti $d$ orbitals. Magnetic properties are very sensitive to electron localization and small changes in partial occupancies, and $d^0$ magnetism is poorly described in local exchange-correlation approximations \cite{droghetti2008,*kovacik2009}. For DFT+$U$, we use the Dudarev {\it et al.} \cite{dudarev1998} implementation. We vary the $U_{\text{eff}}=U-J$ parameter, the difference between local Coulomb repulsion $U$ and exchange interaction $J$, between 0 and 8 eV to study the effect of correlations. Magnetic metals yield best agreement with experiment at $U$ values smaller than those calculated from first principles \cite{petukhov2003}, as the fully localized limit double counting correction exaggerates magnetic moments and band gaps. Therefore, we expect $U_{\text{eff}}$ of 2 to 5 eV (on $p$ orbitals) and 3 to 7 eV (on $d$ orbitals) to yield correct behavior, based on earlier experimental \cite{knotek1978,*ghijsen1988,ray2003} and computational \cite{peng2009,pentcheva2006,dudarev1998,ray2003,gunnarsson1976,*mcmahan1988,*hybertsen1989,*anisimov1991,*mizokawa1995,*solovyev1996,*nekrasov2000,*nandy2010} studies considering $U$ and $J$. Further insight into suitable $U_{\text{eff}}$ is obtained by comparing the magnetization energies with HSEsol calculations, as in Ref. \cite{hong2012}.

\begin{figure}[ht!]
\includegraphics[width=.36\textwidth]{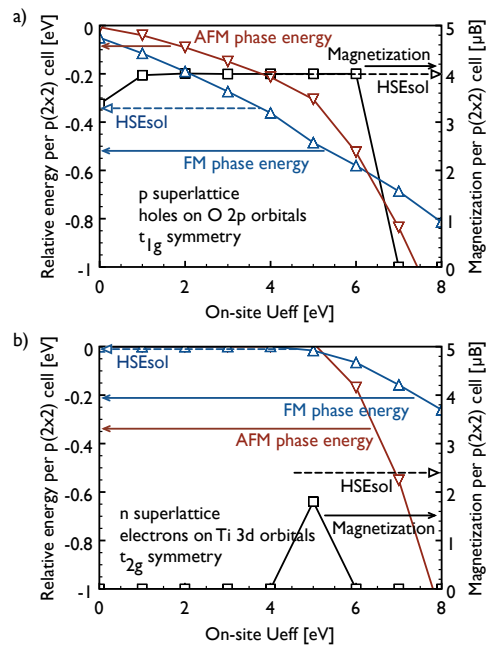}
\caption{\label{fig:magvsu_pandn} (Color online) Calculated a) $p$ type and b) $n$ type \st{}/\kt{} 1.5-1.5 superlattice magnetization (squares, black) and energies of the ferromagnetic (FM, triangles, blue) and antiferromagnetic (AFM, upside down triangles, red) phases relative to paramagnetic (PM) solution, as a function of $U_{\text{eff}}=U-J$ on oxygen $2p$ and Ti/Ta $3d$/$5d$ orbitals. The values are per p(2x2) supercell in GGA+$U$. The lines are guides for the eye. FM phase HSEsol energy and magnetization are marked with dashed arrows.}
\end{figure}

The \st{}/\kt{} superlattices with two $p$ (hole-doped) interfaces are metallic up to $U_{\text{eff}}=6$ eV as well as in HSEsol. The extra holes are confined in the \st{} layers only. The total $p$ (hole-doped) superlattice magnetization and energy difference between ferromagnetic (FM), antiferromagnetic (AFM) and paramagnetic (PM) states, as a function of $U_{\text{eff}}$, are shown in Fig. \ref{fig:magvsu_pandn} a). The DFT+$U$ calculations were done in a $p$(2x2) supercell to find possible antiferromagnetic orderings. The GGA/LSDA system has a partially spin-polarized FM ground state. A small on-site Coulomb interaction considerably increases the energy difference in favor of magnetic ordering, and results in complete spin polarization of the conduction holes with realistic $U_{\text{eff}}$ values (2 to 5 eV). With the HSEsol hybrid functional, as well, half-metallicity is obtained. The HSEsol FM phase energy is indicated in Fig. \ref{fig:magvsu_pandn} a). This would support choosing $U_{\text{eff}}=4$ eV for our system in the spirit of Hong \textit{et al.} \cite{hong2012}.

A charge ordered, symmetry broken AFM solution has 2D AFM ordering of chains of parallel spins. It displays a Mott transition to an insulating state for $U_{\text{eff}}=6$ eV and higher, but it is the ground state only at unphysical $U_{\text{eff}}\gtrsim 7$ eV.


With the $n$ (electron-doped) superlattice, on the other hand, the ground state is paramagnetic metal for small $U_{\text{eff}}$. The electrons are confined in the \st{} layers only. At large $U_{\text{eff}}$ and with the HSEsol hybrid functional, a partially spin-polarized FM state is found, but an AFM phase is energetically preferred (Fig. \ref{fig:magvsu_pandn} b) for almost all $U_{\text{eff}}$. The HSEsol energy difference between the FM and PM phases is only 10 meV per p(2x2) supercell. This would suggest selecting 4 eV $<U_{\text{eff}}<$ 5 eV for $d$ orbitals; however, as seen in Fig. \ref{fig:magvsu_pandn} b), the ground state is highly sensitive to $U_{\text{eff}}$. The energy differences are large only with high $U_{\text{eff}}$, and half-metallicity or Mott insulation are not present. This is in accordance with paramagnetism observed below $d^1$ doping of Ti $d$ orbitals in the bulk \cite{tokura1993}. Complex octahedral rotations \cite{oja2009} and charge and spin ordered phases on the $d$ orbitals \cite{okamoto2006} depending on superlattice geometry, strain and $U_{\text{eff}}$ might be present but are not studied here, because the rotations are not present at room temperature in the constituent perovskites.

Clearly, in the $d^0$ superlattice case, interface holes magnetize much more readily than electrons. To consider the effect of 2D localization on the magnetization of the electron gas, we study a thicker superlattice, where the magnetized hole or electron might spread over 5.5 unit cells of \st{}. The magnitude of magnetization is the same regardless of the thickness of the \st{} layer, demonstrating that ferromagnetism is a true interface effect. The hole density is strongly localized at the interface; in the middle of the 5.5 unit cell \st{} layer, i.e. only two unit cells away from the interface, maximum magnetization density is 20 \% of that at the interface. The strong 2D confinement is comparable to electron gas in $n$ type interfaces \cite{janicka2009}. Coupling of the magnetizations at the two interfaces is negligible, showing that a single interface has a stable FM ground state.




The presence of vacancies can be another source of possible magnetic signals in experiments. Ferromagnetic behavior in oxide materials is often attributed to oxygen vacancies forming localized magnetic moments, as electrons are donated to the surrounding cation $d$ orbitals. In the $n$ type interface, the lack of simple FM coupling in the electron-doped \st{} layer indicates that oxygen vacancies, contributing electrons to the $d$ band, would not cause ferromagnetism. In the $p$ type case, they would counteract hole doping and prevent magnetization. To study the effect vacancies have at interfaces, we calculate a 1.5/1.5 superlattice $p$(2x2) supercell with an electron (or hole) donating O (or K) vacancy. At intermediate $U_{\text{eff}}$ values, $n$ type superlattices with O or K vacancies have close to zero magnetic moment, while $p$ superlattices with O or K vacancies form completely spin-polarized metallic states. The calculated $p$ superlattice magnetic moment is equal to the amount of holes contributed by the interface and the vacancy combined. This means that the interface region is dominated by intrinsic doping, and vacancies will only change the total magnitude of the moment. Metallicity is similarly obtained for 25 \% oxygen vacancy concentration in \st{}/\la{} interfaces \cite{park2006}, indicating that a higher vacancy concentration is needed to change the interface to insulating.



Finally, to find out the effect of the perovskite A and B ions on the localization and magnetization of the hole, we study similar 1.5/1.5 $n$ and $p$ type \st{}/\kn{}, \bt{}/\kn{}, \ct{}/\kn{}, \st{}/\nt{}, \bt{}/\nt{}, and \ct{}/\nt{} superlattices, all of which have one non-polar and one polar $d^0$ perovskite. The obtained behavior of holes and electrons in all structures is identical to the model \st{}/\kt{} system, which demonstrates that ferromagnetism is a general feature of $p$ type interfaces of $d^0$ perovskites. Complete spin polarization of the holes is obtained with small values of $U_{\text{eff}}$, with energies similar to the \st{}/\kt{} case.




To study predicted magnetization, ultrathin epitaxial films of $d^0$ perovskite \kt{}, \kn{}, and \nn{} were grown onto Ti-terminated \st{} (001) single-crystal substrates by \textit{in situ} pulsed laser deposition using a KrF excimer laser and high oxygen pressure (30 Pa during deposition and 10$^5$ Pa during post-deposition cooling). The microstructure of the \kt{}/\st{} interface was characterized by aberration-corrected scanning transmission electron microscopy (STEM) employing the high-angle annular dark field (HAADF) technique. Cross-sectional specimens for STEM analysis were prepared by standard mechanical thinning, precision polishing and Ar ion milling.


\begin{figure}[ht!]
\includegraphics[width=.3\textwidth]{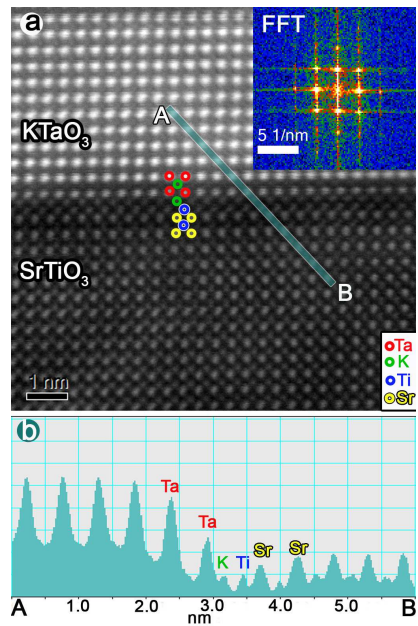}
\caption{\label{fig:stem} (Color online) (a) High-resolution HAADF image of the \kt{}/\st{} interface along the [100] zone axis. The inset shows the local fast Fourier transform pattern. The different elemental atoms close to the interface are marked by colored open circles. (b) Intensity profile of atom columns averaged over the indicated area from A to B in (a).}
\end{figure}

Figure \ref{fig:stem} a)  displays an HAADF image with the viewing direction along the crystallographic [100] direction of the \st{} substrate. From local fast Fourier transform (FFT) analysis (inset of Fig. \ref{fig:stem} (a)), full epitaxial growth of \kt{} on \st{} with a \kt{} [100](001) // \st{} [100](001) relationship is derived. The growth is coherent with the in-plane lattice parameter of the \kt{} film matching that of the \st{} substrate. No dislocations are observed to relax the lattice strain (-2.1 \%) within the first 10 nm of the \kt{} film. In high-resolution HAADF imaging or so-called Z-contrast imaging, the atoms appear with bright contrast on a dark background. An intensity profile across the interface (from A to B) is shown in Fig. \ref{fig:stem} (b). In the \kt{} film, strong peaks from the Ta atoms are easily determined, whereas low intensity peaks from K atoms are not resolved. At the interface, two adjacent low intensity peaks due to Ti and K atoms are visible, indicating the formation of a $p$ type interface. Because the difference in intensity between the Ti and K peaks is below noise level, it is impossible to unambiguously identify the interface structure. While TiO$_2$ termination of the \st{} substrate and the formation of a KO atomic layer at the heterointerface are most likely, some degree of intermixing and off-stoichiometry cannot be excluded.



\begin{figure}[ht!]
\includegraphics[width=.4\textwidth]{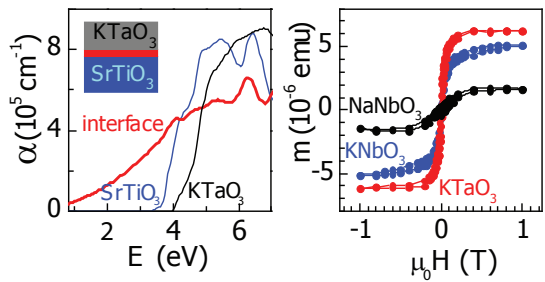}
\caption{\label{fig:temhysteresis} (Color online) a) Absorption coefficient $\alpha$ as a function of photon energy $E$ in \st{} substrate (blue), \kt{} film (black), and interfacial nanolayer (red). The spectra are obtained from ellipsometric data for the model stack shown schematically. b) SQUID measured total magnetization as a function of external magnetic field determined at room temperature in as-deposited 10$\times$5 mm$^2$ films of \kt{}/\st{} (red), \kn{}/\st{} (blue), and \nn{}/\st{} (black).}
\end{figure}

Optical constants in the \kt{}/\st{} heterostructures were determined using variable angle spectroscopic ellipsometry and the WVASE32 software package for data analysis \cite{tyunina2009}. The model analysis considering a stack of substrate, film, surface roughness, and ambient air failed, while a nearly perfect fit was obtained when an additional nanolayer was introduced below the film (Fig. \ref{fig:temhysteresis} a). The thickness of the \kt{} film determined from the ellipsometric data is equal to that determined by x-ray reflectivity, evidencing correctness of the fitting procedure. The thickness of the interfacial layer is about 2 nm. The spectral features of the interfacial nanolayer resemble those of the \st{} substrate at large photon energies $E >$ 3 eV, indicating that the origin of the interfacial layer is related to changes in electronic states in \st{}. A profound absorption tail at $E <$ 3 eV suggests the presence of in-gap states in the interfacial \st{} nanolayer.  

Magnetic measurements were performed in the film plane by a Quantum Design SQUID magnetometer (MPMS XL 7 T). Reciprocating sample transport enabled high sensitivity (10$^{-8}$ emu). The diamagnetic response of the reference \st{} substrate was separately measured and used to extract the magnetization of the film and interface from the response of the film-substrate stack. Additionally, annealing of the samples in an oxygen atmosphere (450-500$^{\circ}$C, 20-24 hours) was performed in order to check the possible influence of oxygen vacancies. Ferromagnetic-type behavior is observed in all samples (Fig. \ref{fig:temhysteresis} b), both as-deposited and after annealing. The saturation moments \footnote{The measurement procedure makes it difficult to accurately determine the magnitude of the rather small saturation moments. However, it allows for a comparison of the magnetization in the same sample immediately after deposition with that after annealing and a study of magnetization as a function of temperature.} do not decrease after annealing and they remain practically unchanged with decreasing temperature from 300 to 20 K, which indicates ordering temperatures well above room temperature.

\begin{figure}[ht!]
\includegraphics[width=.49\textwidth]{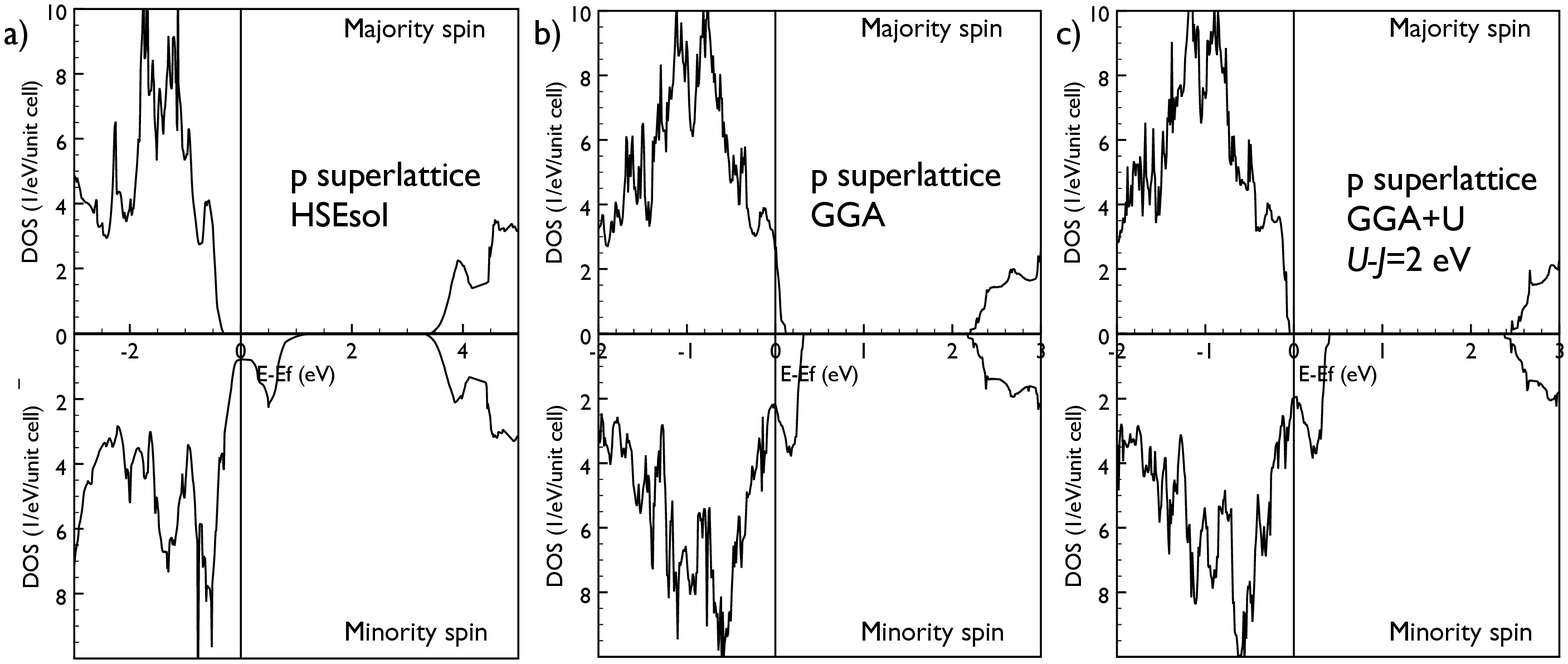}
\caption{\label{fig:dos_combined} Densities of states for 1.5/1.5 \st{}/\kt{} p type superlattice with a) HSEsol hybrid functional, b) GGA and c) GGA+$U$, $U_{\text{eff}}$=2 eV.}
\end{figure}

As stated, in $d^0$ bulk perovskites and superlattices, the $d$ band is very wide, and the oxygen $2p$ band has a higher DOS and the exchange interaction $J$ is higher for oxygen $p$ than transition metal $d$ states \cite{peng2009}. The HSEsol hybrid functional yields a half-metallic density of states where the $2p$ band edges of the two spins are separated by over 1 eV in the $p$ interface case (Fig. \ref{fig:dos_combined} a)). For the $n$ type interface FM phase (not pictured), on the other hand, the HSEsol Ti $d$ band edge stays almost the same for both spins, although exchange splitting of the peak occurs. The strength of magnetic effects in these interfaces is illustrated by the HSEsol magnetization energy, which is 340 meV for $p$ type interface but only 10 meV for $n$ type interface per p(2x2) supercell.

In the Stoner ferromagnetic transition, the spike in the DOS crosses the Fermi level when spin polarization takes place. This also happens in the LSDA and GGA DOS (Fig. \ref{fig:dos_combined} b)), and the system seems to be close to a fully half-metallic state. Half-metallic FM ground state requires a small local Coulomb repulsion (Fig. \ref{fig:dos_combined} c) or inclusion of nonlocal exchange (Fig. \ref{fig:dos_combined} a) on the oxygen $p$ orbitals. The electron-doped superlattice with small $U_{\text{eff}}$, on the other hand, retains a paramagnetic DOS virtually identical to the GGA DOS, and nonlocal exchange causes only partial magnetization. The hole-doped system clearly satisfies the Stoner criterion $JD(E_f)>1$, while the electron-doped does not. The same mechanism has been suggested for hole-doped ZnO \cite{peng2009}, where non-intrinsic doping methods have been considered.


In summary, we report experimental evidence and \textit{ab-initio} analysis of $d^0$ ferromagnetism at charge-imbalanced perovskite interfaces. We suggest that ferromagnetic ordering and half-metallicity is present in hole-doped $d^0$ perovskites due to large $p$ orbital exchange splitting and the high density of states at the top of the oxygen $p$ band under sufficient doping. The reason for the general scarcity of $d^0$ ferromagnets is the difficulty of doping the $p$ valence band. High hole concentrations, however, are easily obtained with intrinsic doping by interfaces. Even a single interface has sufficient DOS for ferromagnetism, since the carriers are strongly 2D localized at the interface region. Electron-doped $d^0$ perovskites are ordinary metals, as the bottom of the cation $d$ band has larger dispersion, but they are susceptible to partial magnetization and possible complex magnetic ordering patterns under 2D localization.

We acknowledge useful discussions with Javad Hashemi, Ville Havu and Torbj\"orn Bj\"orkman. CSC (the Finnish IT Center for Science) provided the computing resources. The work has been supported by the Academy of Finland through its Centers of Excellence Program (project no. 251748) and the FinNano Program (project no. 128229), and a grant from the V\"ais\"al\"a Fund of the Finnish Academy of Science and Letters.


%

\end{document}